\documentclass[12pt]{article}
\usepackage{a4wide}
\usepackage{axodraw}
\usepackage{amsmath}
\usepackage{fancybox}
\newcommand{\half}{{\textstyle\frac12}}
\newcommand{\mbar}{\overline{m}_b}
\newcommand{\msbar}{\overline{\textrm{\small{MS}}}}
\newcommand{\GeV}{\mathrm{GeV}}
\begin{document}
% start of shep titlepage
\begin{flushright}
ROME1-1234/98\\ 
SHEP 98/14\\
hep-lat/9812001
\end{flushright}
\vspace{3em}

\begin{center}
{\Large\bfseries Computation of the $\mathbf{b}$-Quark Mass}\\ 
{\Large\bfseries with Perturbative Matching at the
Next-to-Next-to-Leading Order}
\\[2em] G.~Martinelli$^1$ and C.T.~Sachrajda$^2$\\[1em] $^1$
Dip. di Fisica, Univ. ``La Sapienza'' and\\ INFN, Sezione di Roma, Ple
A. Moro, I-00185, Rome, Italy\\
\vspace{10pt}
$^2$ Department of Physics and Astronomy, University of Southampton\\
Southampton SO17 1BJ, UK
\end{center}
\vspace{1in}

\begin{center}\textbf{Abstract}\end{center}
\begin{quote}
We compute the two-loop term in the perturbation series for the
quark-mass in the lattice Heavy Quark Effective Theory (HQET). This is
an ingredient in the matching factor required to obtain the $b$-quark
mass from lattice simulations of the HQET.  Combining our calculation
with numerical results from the APE collaboration, we find, at
two-loop order, $\mbar\equiv m_b^{\overline{\textrm
MS}}(m_b^{\overline{\textrm MS}})=4.41\pm0.05\pm0.10$\,GeV. It was
expected that the two-loop term would have a significant effect, and
this is indeed what we find.  Depending on the choice of ``reasonable"
coupling constant in the one-loop estimates, the result for $\mbar$
can change by several hundred MeV when the two-loop terms are
included.
\end{quote}
\newpage

\section{Introduction}

In this paper we evaluate the two-loop perturbative matching factor
required to obtain the mass of a heavy quark from lattice simulations of
the Heavy Quark Effective Theory (i.e. from simulations of static heavy
quarks). This is the next-to-next-to-leading-order (NNLO) term in the
matching factor. We combine this matching factor with numerical results
from simulations to obtain:
\begin{equation}
\Ovalbox{$\rule[-.25cm]{0cm}{.7cm}
\ \ \mbar = 4.41\pm 0.05 \pm 0.10\ \mathrm{GeV}$\ \ } 
\label{eq:mbarresult}\end{equation}
where
\begin{equation}
\overline{m}_b\equiv m_b^{\overline{\textrm
MS}}(m_b^{\overline{\textrm MS}})\ ,
\label{eq:mbardef}\end{equation}
at NNLO. The first error in eq.~(\ref{eq:mbarresult}) is due to the
uncertainties in the values of the lattice spacing, in the value of the
strong coupling constant and in the numerical evaluation of the
functional integral. The second error is an estimate of the
uncertainties due to  our ignorance of 3-loop and higher order
perturbative terms.

For the Lagrangian of the HQET we take:
\begin{equation}
{\cal L}_{\mathrm{HQET}}=\bar h D_4 \frac{1+\gamma^4}{2}h\ ,
\label{eq:lh}\end{equation}
where $h$ is the field of the heavy quark, and we use the following
definition of the covariant derivative $D$: 
\begin{equation}
D_\mu\,h(x)=U_\mu(x)h(x+\hat\mu)-h(x)\ ,
\label{eq:dmudef}\end{equation}
where $x+\hat\mu$ is the neighbouring point to $x$ in the $\mu$-direction.
$U_\mu(x)$ is the link variable from $x$ to $x+\hat\mu$. For the
light-quark action we take an ``improved'' generalisation of the
Wilson action. For each quark flavour the action is:
\begin{equation}
S_q=S_W +  c_{\mathrm{SW}}\,S_{\mathrm{SW}}\ ,
\label{eq:sq}\end{equation}
where ${\cal S}_W$ is the Wilson fermion action:
\begin{multline}
S_W=\sum_x\Bigg\{-\frac12\sum_\mu\Big[
\overline\psi(x)(1-\gamma_\mu)U_\mu(x)\psi(x+\hat\mu)
+\overline\psi(x+\hat\mu)(1+\gamma_\mu)U_\mu^\dagger(x)\psi(x)
\Big]\\ +\overline\psi(x)(m_0+4)\psi(x)\Bigg\}\ ,\hspace{.5in}
\label{eq:wilsonaction}\end{multline}
and $S_{\mathrm{SW}}$ is the Sheihkoleslami-Wohlert (or ``clover'')
action~\cite{sw}
\begin{equation}
S_{\mathrm{SW}}=-\frac{i}{4}\sum_{x,\mu,\nu}\Big[\overline\psi(x)
\sigma_{\mu\nu}F_{\mu\nu}(x)\psi(x)\Big]\ .
\label{eq:swaction}\end{equation}
$F_{\mu\nu}$ is the lattice expression of the field-strength tensor
obtained by averaging the four plaquettes lying in the $(\mu,\nu)$
plane and stemming from the point $x$. The coefficient
$c_{\mathrm{SW}}$ is equal to 1 at tree-level (with
$c_{\mathrm{SW}}=1$ and using appropriate operators the discretisation
errors in physical quantities are reduced from $O(a)$ for the Wilson
action to $O(\alpha_s\,a)$). Recently much effort has been devoted to
fixing $c_{\mathrm{SW}}$ non-perturbatively in such a way that the
errors are reduced to ones of $O(a^2)$~\cite{luschercsw}. For the
perturbative calculations in this paper we will keep $c_{\mathrm{SW}}$
and $N_f$, the number of light-quark flavours as free 
parameters~\footnote{The results are also presented
for an arbitrary number of coulours, $N$.}.

\begin{figure}
\begin{center}
\begin{picture}(350,60)(-50,10)
\SetWidth{1.5}\Line(-50,20)(90,20)\SetWidth{0.5}
\GlueArc(20,20)(40,0,180){2}{20}
\Text(20,10)[t]{(a)}
\SetWidth{1.5}\Line(180,20)(320,20)\SetWidth{0.5}
\GlueArc(235,20)(30,0,180){2}{15}
\GlueArc(265,20)(30,0,180){2}{15}
\Text(250,10)[t]{(b)}
\end{picture}
\end{center}
\caption{(a) A one-loop diagram and (b) a two-loop diagram
contributing to $\delta m$. The solid lines represent static
heavy quarks.}\label{fig:deltam} 
\end{figure}
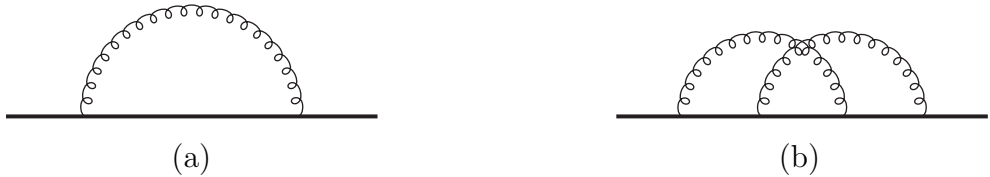

The starting point for our evaluation of the heavy quark mass is
the computation of the correlation functions of the time-component
of two axial currents:
\begin{eqnarray}
C(t) & = & \sum_{\vec x} \langle\,0\,|A_4(x)\,A^\dagger_4(0)|\,0\,\rangle
\\ 
& \simeq&Z^2\,\exp(-{\cal E}t)\ ,
\end{eqnarray}
where we have assumed that $t$ is positive and sufficiently large so
that we can neglect the contributions from excited states. The current
takes the form~\footnote{In practice it is generally advantageous to
use ``smeared'' interpolating operators in order to enhance the
overlap with the ground state.} $A_4(x)=\overline
h(x)\gamma_4\gamma_5q(x)$ (the factor of $\gamma_4$ can be replaced by
the Identity since it is adjacent to the heavy quark field). By
fitting the correlator $C(t)$ as a function of $t$, both the prefactor
$Z$ and the exponent $\cal E$ can be determined. From $Z$ we can
obtain the decay constant of a heavy pseudoscalar meson in the static
limit, whereas from $\cal E$ we obtain
$\mbar$~\cite{crisafulli}. Matching QCD and the Lattice HQET one finds
that~\cite{crisafulli}:
\begin{equation}
m_b^{\textrm{pole}}=M_B-{\cal E} + \delta m\ ,
\label{eq:mbpole}\end{equation}
where $m_b^{\textrm{pole}}$ is the ``pole'' mass of the $b$-quark, $M_B$
is the physical mass of the $B$-meson and $\delta m$ is the mass
generated in the static theory, eq.~(\ref{eq:lh}), in perturbation
theory ($a\delta m = \alpha_s\,X_0 + \alpha_s^2\,X_1$, where the
coefficient $X_1$ depends on the choice of coupling constant,
$\alpha_s$, used in the perturbative expansion). Even though there is no
explicit mass-term in the bare action of eq.~(\ref{eq:lh}), Feynman
diagrams, such as those in fig.~\ref{fig:deltam}, generate such a term
which is proportional to $1/a$, where $a$ is the lattice spacing. Once
$m_b^{\textrm{pole}}$ is known, then any physical, short-distance, quark
mass can be determined using continuum perturbation theory, e.g. 
\begin{equation}
\overline{m}_b=\left(M_B-{\cal E}+\delta m\right)\,\left[1-\frac43\left(
\frac{\alpha_s(\mbar)}{\pi}\right)-(11.66-1.04N_f)
\left(\frac{\alpha_s(\mbar)}{\pi}\right)^2\,+O\left(\alpha^3_s(\mbar)
\right)\right],
\label{eq:key}\end{equation}
where $N_f$ is the number of light-quark flavours and $\alpha_s$ is
defined in the $\msbar$ renormalization scheme. The term in square
brackets in eq.~(\ref{eq:key}) is the pertubative expansion of the
factor which relates the pole mass to the $\msbar$
one~\cite{broadhurst}. \textit{The main aim of this paper is to
calculate $\mathit{\delta m}$ to two-loop order}. Using the $\msbar$
coupling $\alpha_s(\mbar)$ as the expansion parameter we find 
\begin{multline}
a\,\delta m  =  2.1173\,\alpha_s(\mbar)+
\{\,(3.707-0.225\,N_f)\ln(\mbar a)\\
\hspace{1in}-1.306 -N_f\,(0.104+0.100\,c_{\mathrm{SW}}
-0.403\,c^2_{\mathrm{SW}})\,\}\,\alpha^2_s(\mbar)\ .
\label{eq:deltaunquenched0}\end{multline}

Consider eqs.~(\ref{eq:mbpole}), (\ref{eq:key}) and
(\ref{eq:deltaunquenched0}). Neither the perturbation series for $\delta
m$ nor that relating $m_b^{\textrm{pole}}$ and $\mbar$ in the square
brackets in eq.~(\ref{eq:key}) is Borel summable and both contain
renormalon ambiguities. In each case the leading ambiguity is of
$O(\Lambda_{\textrm QCD})$. Since $m_b^{\textrm{pole}}$ is not a
physical quantity, the presence of a renormalon on the right-hand side
of eq.~(\ref{eq:mbpole}) is not inconsistent. $\mbar$, however, is a
physical quantity, and the leading renormalon ambiguity cancels when the
two perturbation series in eq.~(\ref{eq:key}) are
combined~\cite{bb,bsuv,ht}~\footnote{In order for this cancellation to
be manifest, a single coupling constant for both perturbation series
must be used (we use $\alpha_s(\mbar)$).}. The remaining renormalon
ambiguity is of $O(\Lambda_{\textrm{QCD}}^2/\mbar)$ which is beyond the
precision we are considering in this paper. Related to the problem of
renormalons is that of power divergences (i.e. the terms which behave as
powers of $a^{-1}$). $\delta m$ diverges linearly with the inverse
lattice spacing, cancelling the linear divergence in ${\cal E}$ (this
cancellation is necessarily only partial since we truncate the
perturbation series for $\delta m$ at a finite order of perturbation
theory). Thus by computing ${\cal E}$ numerically, and calculating
$\delta m$ and the relation between the pole-mass and $\mbar$ in
perturbation theory (this relation is known to two-loop
order~\cite{broadhurst}), $\mbar$ can be obtained. It is free of power
divergences and renormalon ambiguities.

Although we do not present the discussion explicitly in the formalism
of an Operator Product Expansion, the calculation of the quark mass
here is an example of the evaluation of a power correction. To lowest
order in the heavy quark expansion we can take $m_b=M_B$, and we are
evaluating the first power corrections to this (i.e. the correction of
$O(\Lambda_{\textrm{QCD}})$ relative to the leading term). In
ref.~\cite{ht} we argued that the evaluation of power corrections is
difficult, requiring high-orders of perturbation theory in order to
subtract the power divergences to sufficient precision. To illustrate
this point for the heavy-quark mass, we show in
subsection~\ref{subsubsec:oneloop} that the one-loop result changes by
several hundred MeV depending on which ``reasonable" coupling constant
is used in the one-loop result (specifically we use the $\msbar$
coupling at an average momentum $q^*$, the $V$ coupling defined from
the potential also at the scale $q^*$ and a boosted lattice coupling
constant). The two-loop calculation described in this paper reduces
this uncertainty to 100~MeV or less.

In the present calculation the ultra-violet divergences are linear in
the inverse lattice spacing. For other physical quantities for which the
power divergences are quadratic (or even higher order) it is even more
difficult to control the perturbative corrections~\cite{ht}.

The plan for the remainder of this paper is as follows. In the next
section we present the outline of the perturbative calculation of
$\delta m$ up to two-loop order. In an attempt to make the steps of
the calculation clearer, we have relegated the technical details to a
(rather long) appendix. In section~\ref{sec:mb} we use our
perturbative results in order to determine the mass of the $b$-quark
from values of $\cal E$ obtained in numerical simulations. Finally in
section~\ref{sec:concs} we present our conclusions.

\section{Calculation of $\delta m$ to Two-Loop Order} In this section we
outline the evaluation of $\delta m$ up to two-loop order in
perturbation theory. Instead of evaluating self-energy diagrams, such as
those in fig.~\ref{fig:deltam}, directly, we exploit the fact that
two-loop perturbative results for rectangular Wilson loops have already
been presented~\cite{hk,curci}. We therefore determine $\delta m$ from
the exponential behaviour of large Wilson loops as we now explain. 

Consider a Wilson loop (W(R,T)) in the $\mu$-$\nu$ plane, of length
$R$ and $T$ in the $\mu$ and $\nu$ directions respectively. Since,
\textit{in perturbation theory}, the potential between static-quarks
falls like $1/r$ with their separation $r$, the expectation value of
large Wilson loops decrease exponentially with the perimeter of the
loops, specifically
\begin{equation}
\langle\ W(R,T)\ \rangle\,\sim\,\exp(-2\delta m(R+T))\ .
\label{eq:wasymp}\end{equation}
Hence the perimeter term in $\log(\langle\,W(R,T)\,\rangle)$ is simply
$-2\delta m(R+T)$.

Following ref.~\cite{hk} we define the first two coefficients of the
perturbative expansion for $\langle\,W(R,T)\,\rangle$ as follows:
\begin{equation}
\langle\ W(R,T)\ \rangle=1-g^2\,W_2(R,T)-g^4\,W_4(R,T) +O(g^6)\ ,
\label{eq:wexp}\end{equation}
where $g$ is the bare lattice coupling constant.
By using eq.~(\ref{eq:wasymp}) we obtain the perturbative expansion
for $\delta m$ in terms of these coefficients,
\begin{equation}
\delta m = \frac{1}{2(R+T)}\Big[g^2\,W_2(R,T)+ g^4\big(W_4(R,T)+\half
W_2^2(R,T)\big)\Big]\ .
\label{eq:deltamexp}\end{equation}
We now evaluate $W_2$ and $W_4$, only keeping terms which grow at least
linearly with the dimensions of the Wilson loop.

\subsection{Evaluation of $\mathbf{W_2(R,T)}$}
\label{subsec:w2}
In ref.~\cite{hk} we find the following integral representation of
$W_2$ for an $SU(N)$-gauge theory:
\begin{equation}
W_2(R,T)=\frac{(N^2-1)}{N}
\int_{-\pi}^\pi\frac{d^4p}{(2\pi)^4}\ 
\frac{\sin^2(\half p_\nu T)
\sin^2(\half p_\mu R)}{D(p)}\left\{
\frac{1}{\sin^2(\half p_\mu)}+\frac{1}{\sin^2(\half p_\nu)}\right\}\ ,
\label{eq:w2int}\end{equation}
where
\begin{equation}
D(p)=4\sum_{\lambda=1}^4\sin^2(\half p_\lambda)\ .
\label{eq:ddef}\end{equation}
Consider the first term in the integrand in eq.~(\ref{eq:w2int}), i.e.
the term containing a factor of $1/\sin^2(\half p_\mu)$. By inspection
(i.e. by using power counting) it can be seen that the region of small
$p_\mu$ gives a contribution proportional to $R$ (and similarly the
second term gives a contribution proportional to $T$). By changing
variables from $p_\mu$ to $z=\exp(ip_\mu)$ (so that the contour of
integration in $z$ is the unit circle), and studying the singularities
of the integrand in $z$, the integration over $z$ can be performed. In
this way one obtains
\begin{equation}
W_2(R,T)\equiv\frac{(N^2-1)}{4N}\overline W_2(R,T)
=\frac{(N^2-1)}{4N}\big\{(R+T)\,X - Y(R) - Y(T)\big\}\ ,
\label{eq:w2res}\end{equation}
where $X$ and $Y$ are three-dimensional integrals defined in
eqs.~(\ref{eq:xdef}) and (\ref{eq:ydef}) of the appendix. $X$ is a
finite integral, whereas $Y(R)$ grows logarithmically with
$R$. $\delta m$ at one-loop level is given by the term in the
perimeter which is proportional to $X$.  We will, however, require
the terms proportional to $Y(R)+Y(T)$ when we evaluate $\delta m$ at
two-loops (see eq.~(\ref{eq:deltamexp})\,).  Equation~(\ref{eq:w2res})
implies that
\begin{equation}
\delta m = \frac{N^2-1}{8N}\,X\,g^2\ + \ O(g^4)\ ,
\label{eq:deltam1loopg}\end{equation}
which is the well known one-loop result.

\subsection{Evaluation of $\mathbf{W_4(R,T)}$}
The evaluation of $W_4(R,T)$ is the main calculation of this paper, and
we describe this calculation in some detail in the Appendix. We write 
\begin{equation}
W_4(R,T)=W_4^{\mathrm{gluon}}(R,T) +
W_4^{\mathrm{fermion}}(R,T)\ ,
\label{eq:w4res}\end{equation}
where
$W_4^{\mathrm{gluon}}(R,T)$ is the contribution from diagrams in the
pure-gauge theory and \\ $W_4^{\mathrm{fermion}}(R,T)$ is that from
diagrams containing light-quark loops. The purely gluonic contribution
is given in eq.~(\ref{eq:w4gluonres})
\begin{multline}
W_4^{\mathrm{gluon}}(R,T)=-\frac{(N^2-1)^2}{32N^2}X^2(R+T)^2 
+\frac{(N^2-1)^2}{16N^2}X\,(Y(R)+Y(T))\,(R+T)\\
+\frac{(N^2-1)}{192}\left\{3X^2 - 2XL+ 12W\right\}\,(R+T)\\ 
+\frac{(N^2-1)}{96}\left\{3V_1 + 6V_2 + 7LX -
\frac{5}{4}X\right\}\,(R+T)  + \left\{\frac{(N^2-1)(2N^2-3)}
{96N^2}X\right\}(R+T)\ ,
\label{eq:w4gluonres0}\end{multline}
and the contribution from graphs containing light-quark loops is given
in eq.~(\ref{eq:w4fermionres})
\begin{equation}
W_4^{\mathrm{fermion}}(R,T)=\frac{(N^2-1)N_f}{16N}\,(R+T)\,
\big(V_{\mathrm{W}}+V_{\mathrm{SW}}\big)\ ,
\label{eq:w4fermionres0}\end{equation}
where $N_f$ is the number of active light-quark flavours. The integrals
$X,L,Y,W,V_{1,2},V_{\textrm{W}}$ and $V_{\textrm{SW}}$ in these
expressions are defined and evaluated in section~\ref{subsec:integrals}
of the appendix. $V_{\textrm{SW}}$ is the contribution from the
improvement term in the fermionic action and we write it as 
$V_{\textrm{SW}}=V^a_{\textrm{SW}}\,c_{\textrm{SW}} +
V^b_{\textrm{SW}}\,c^2_{\textrm{SW}}$. With the exception of $Y(R)$,
which depends logarithmically on $R$ and cancels out of the expression
for $\delta m$, we present the numerical values of these integrals in
table~\ref{tab:xlwv}.

\begin{table}
\begin{center}
\begin{tabular}{|c|c|c|}\hline
Integral & Defined in eq. & Value \\ \hline
$X$ & (\ref{eq:xdef}) & 0.50546 \\
$L$ & (\ref{eq:ldef}) & 0.30987 \\ 
$W$ & (\ref{eq:wdef}) & $1.313(3)\cdot 10^{-2}$\\ 
$V_1$ & (\ref{eq:v1def}) & $7.008(5)\cdot 10^{-2}$\\
$V_2$ & (\ref{eq:v2def}) & $-2.130(1)\cdot 10^{-2}\ \,$\\
$V_{\textrm{W}}$ & (\ref{eq:vwdef}) & $-2.145(1)\cdot 10^{-2}\,\ $\\
$V^a_{\textrm{SW}}$ & (\ref{eq:vswdef}) & $\ \,\,2.63(2)\cdot 10^{-3}$\\
$V^b_{\textrm{SW}}$ & (\ref{eq:vswdef}) & $-2.974(1)\cdot 10^{-2}\,\ $\\
\hline
\end{tabular}
\caption{Values of the integrals used in the evaluation of $\delta m$ to
two-loops in perturbation theory.\label{tab:xlwv}}
\end{center}
\end{table}

\subsection{The Perturbation Series for $\mathbf{\delta m}$}
Having determined the relevant terms in $W_2(R,T)$ and $W_4(R,T)$ we
can readily obtain the perturbation series for $\delta m$ using
eq.~(\ref{eq:deltamexp}). At one-loop order, the result is presented
in eq.~(\ref{eq:deltam1loopg}) which, setting the number of colours
$N$ equal to 3, we rewrite as:
\begin{equation}
a\,\delta m \simeq 2.1173\,\alpha_0 +O(\alpha_0^2)\ ,
\label{eq:deltam1loop}\end{equation}
where the subscript $0$ on $\alpha_0$ reminds us that the expansion is
in terms of the bare coupling constant in the lattice theory.

The two-loop contribution to $\delta m$ is proportional to $W_4(R,T)
+\half W_2^2(R,T)$ (see eq.~(\ref{eq:deltamexp})\,): 
\begin{multline}
W_4(R,T) +\half W_2^2(R,T) = (R+T)\frac{(N^2-1)}{192}
\, \Bigg\{3X^2 + 12W
+6V_1 + 12V_2 + \\ 12 LX 
+ \frac{3(N^2-4)}{2N^2}\,X+\frac{12 N_f}{N}(V_{\mathrm{W}}+
V_{\mathrm{SW}})\Bigg\}\ .
\label{eq:w4w2sq}\end{multline}
The terms quadratic in $R$ and $T$ explicitly cancel in the combination
$W_4+\half W_2^2$, as do those proportional to the integrals $Y(R)$  and
$Y(T)$ as they must. This is an important consistency check of our
calculations. All the integrals on the right-hand side of
eq.~(\ref{eq:w4w2sq}) are finite and independent of $R$ and $T$.
Substituting $N=3$ and the numerical values of the integrals (see the
appendix) in the right-hand side of eq.~(\ref{eq:w4w2sq}), we
obtain~\footnote{The errors in the coefficients in
eq.~(\ref{eq:logwres}) (and following equations) are due to those in the
numerical values of the integrals given in the appendix. They are
typically 1 (or possibly 2) in the last digit, and are negligible in the
evaluation of $\mbar$.}
\begin{equation}
W_4(R,T)\hspace{-1.5pt}+\hspace{-1.5pt}\half W_2^2(R,T)
\hspace{-1.5pt} = \hspace{-1.5pt} (R+T)\,\{0.14124 + N_f(-0.00358
+0.00044\,c_{\mathrm{SW}}-0.00496\,c_{\mathrm{SW}}^2)\}.
\label{eq:logwres}\end{equation}
We thus arrive at the two-loop expression for $\delta m$:
\begin{equation}
\Ovalbox{$\rule[-.3cm]{0cm}{.8cm}\ a\,\delta m 
\simeq  2.1173\,\alpha_0 +\{11.152 + N_f(-0.282+0.035\,c_{\mathrm{SW}}
-0.391\,c_{\mathrm{SW}}^2)\}\,\alpha_0^2
+O(\alpha_0^3)\ .\ $}
\label{eq:deltam2loop}\end{equation}
Eq.~(\ref{eq:deltam2loop}) is the main result of this paper. In the
following section we will exploit this result to extract the mass of
the $b$-quark from the values of $\cal E$ computed in lattice
simulations.

\section{Lattice Computation of $\mathbf{m_b}$}
\label{sec:mb}
In this section we discuss the evaluation of the heavy quark mass,
from lattice computations of $\cal E$ using the two-loop result for
$\delta m$ in eq.~(\ref{eq:deltam2loop}). For illustration we will
determine the $\overline{\textrm{\small{MS}}}$-mass, $\mbar$, defined
in eq.~(\ref{eq:mbardef}), from which any other short-distance
definition of the heavy-quark mass can be obtained using continuum
perturbation theory.

As discussed in the introduction, the relation between
$\overline{m}_b$ and ${\cal E}$ is:
\begin{equation}
\overline{m}_b=\left(M_B-{\cal E}+\delta m\right)\,\left[1-\frac43\left(
\frac{\alpha_s(\mbar)}{\pi}\right)-(11.66-1.04N_f)
\left(\frac{\alpha_s(\mbar)}{\pi}\right)^2\right]\ ,
\label{eq:key0}\end{equation}
where $N_f$ is the number of light-quark flavours and $\alpha_s$ is
defined in the $\msbar$ renormalization scheme. In eq.~(\ref{eq:key}),
$M_B=5.279$~GeV is the physical mass of the $B$-meson, ${\cal E}$ is
computed in lattice simulations and the remaining terms are calculated
in perturbation theory. The term in square brackets in
eq.~(\ref{eq:key}) is the perturbation expansion of the factor which
relates the pole mass to the $\msbar$ one~\cite{broadhurst}.

The two-loop expression for $\delta m$ in eq.~(\ref{eq:deltam2loop}) is
given in terms of $\alpha_0$, the bare coupling constant in the lattice
theory. In order to achieve the explicit cancellation of renormalon
singularities the same coupling constant needs to be used in the
expansion of $\delta m$ and in the relation between the pole mass and
the $\msbar$ mass (i.e. in the expression in square parentheses in
eq.~(\ref{eq:key})\,). The relation between the $\msbar$ coupling
$\alpha_s$ and the lattice coupling $\alpha_0$ is:
\begin{equation}
\alpha_s\left(\frac{s}{a}\right)=\alpha_0+d_1(s)\,\alpha_0^2 +
d_2(s)\,\alpha_0^3 + \cdots\ ,
\label{eq:d1def}\end{equation}
where
\begin{equation}
d_1(s)=-\frac{\beta_0}{2\pi}\ln(s) -\frac{\pi}{2N}+2.13573N
+N_f\,\left(-0.08413 + 0.0634\,c_{\mathrm{SW}} -
0.3750\,c_{\mathrm{SW}}^2\right)\ ,
\label{eq:d1res}\end{equation}
and $\beta_0=(11N-2N_f)/3$. The contribution to the right-hand side of
eq.~(\ref{eq:d1res}) for the gauge theory without fermions is found in
ref.~\cite{hasenfratz}, and the quark contribution  for Wilson fermions
(i.e. with $c_{\mathrm{SW}}=0$) can be obtained from ref.~\cite{kawai}.
The coefficients of $c_{\mathrm{SW}}$ and $c^2_{\mathrm{SW}}$ are
presented here for the first time.  Although not needed for our present
calculation, the two-loop coefficient in eq.~(\ref{eq:d1def}), $d_2(s)$,
for the pure-gauge theory can be found in ref.~\cite{luscher}.

Combining eqs.~(\ref{eq:deltam2loop}) and (\ref{eq:d1res}) we obtain the
result in eq.~(\ref{eq:deltaunquenched0}) for $\delta m$ expressed in
terms of the $\msbar$ coupling.

\subsection{Numerical Results}
The results for $\cal E$ computed in lattice simulations to date, have
been obtained in the quenched approximation. For the purposes of this
study we take the results from the APE collaboration~\cite{ape}:
\begin{eqnarray}
a{\cal E}& =& 0.61(1) \ \ \ \textrm{at\ }\ \beta=6.0\ \ \ 
(a^{-1}=2.0\,(2)\,\GeV)\label{eq:cale60}\\ 
a{\cal E}& =& 0.52(1) \ \ \ \textrm{at\ }\ \beta=6.2\ \ \ 
(a^{-1}=2.9\,(3)\,\GeV)\label{eq:cale62}\\ 
a{\cal E}& =&0.460(7) \ \ \textrm{at\ }\ \beta=6.4\ \ \ 
(a^{-1}=3.8\,(3)\,\GeV)\ .
\label{eq:cale64}\end{eqnarray}
Of course the quenched calculation is incomplete and there is no
procedure for determining $\mbar$ from quenched computations of $\cal
E$ which is totally satisfactory. We will now describe our approach,
but note that many readers may choose to follow different procedures,
which are equally valid in the absence of unquenched results. We take
the quenched results in eqs.~(\ref{eq:cale60})--(\ref{eq:cale64}) and
determine the pole mass by combining them with the perturbative result
for $\delta m$ written in terms of the $\msbar$ coupling constant at
the scale $\mbar$ obtained using eqs.~(\ref{eq:d1def}) and
(\ref{eq:d1res}) with $N_f=0$. We then derive $\mbar$ from this value
of the pole mass using continuum perturbation theory,
eq.~(\ref{eq:key}), but also with $N_f=0$, to ensure the cancellation
of renormalon singularities. We present the details and results in
subsection~\ref{subsec:quenched}.

In subsection~\ref{subsubsec:unquenched} below we discuss the
fermionic contributions in perturbation theory and demonstrate that
they are large (of $O(100)$\,MeV or more for the lattice spacing
considered in this paper). Finally in this section we compare our
two-loop results with some standard procedures used to try to optimise
the one-loop results, such as using the boosted coupling constant or a
coupling defined at some average momentum for the process, $q^*$, (see
subsection \ref{subsubsec:oneloop}).

\subsubsection{Quenched Results for $\mathbf{\mbar}$} 
\label{subsec:quenched}
In the quenched approximation we can rewrite $\delta m$ in the form
\begin{equation}
a\,\delta m = 2.1173\,\alpha_s(\mbar)+
(3.707\ln(\mbar a)-1.306)\,\alpha^2_s(\mbar)\ .
\label{eq:deltanew}\end{equation}
Most of the very large coefficient 11.152 in the perturbation series
written in terms of $\alpha_0$ (see eq.~(\ref{eq:deltam2loop})\,) has
been absorbed into the coupling $\alpha_s$. This is frequently the
case~\cite{lm}. The implicit equation for $\mbar$ is (see
eq.~(\ref{eq:key})\,)
\begin{multline}
\mbar=\Delta\,\left[1+\frac{1}{\Delta a}\left\{2.1173\,\alpha_s(\mbar)
+(3.707\ln(\mbar a)-1.306)\,\alpha^2_s(\mbar)\right\}\right]\\ 
\times\left[1-\frac43\frac{\alpha_s(\mbar)}{\pi}
-11.66\,\left(\frac{\alpha_s(\mbar)}
{\pi}\right)^2\right]\hspace{0.5in}
\label{eq:mbar1}\end{multline}
where $\Delta=m_B-{\cal E}$. To the order we are working in,
eq.~(\ref{eq:mbar1}) is equivalent to:
\begin{multline}
\mbar=\Delta\,\left[1 + \left\{\frac{2.1173}{\Delta a}-0.4244\right\}\alpha_s(\mbar)
\right.
\\\left.
+\left\{\frac{1}{\Delta a}\left(3.707\ln(\mbar a)-2.207\right)-
1.181\right\}
\,\alpha^2_s(\mbar)+\dots\right]\ .
\label{eq:mbar2}\end{multline}
The difference of the results obtained using eqs.~(\ref{eq:mbar1}) and
(\ref{eq:mbar2}) will be an indication of the error due to 3-loop and
higher order terms.

We estimate $\mbar$ using equations (\ref{eq:mbar1}) and
(\ref{eq:mbar2}). At this stage, in spite of using the quenched results
for $\cal E$, we simply assume that the result is the physical one and 
for $\alpha_s(\mbar)$ we take the coupling  constant obtained from
$\alpha_s(M_Z)=0.118(3)$ using the two-loop evolution equation with five
flavours (this gives $\alpha_s(\mbar)\simeq 0.22(1)$\,).
Eqs.~(\ref{eq:mbar1}) and (\ref{eq:mbar2}), which contain $\mbar$ both
on the left and right-hand sides, are solved by iteration and the
corresponding results are presented in table~\ref{tab:mbar}.
\begin{table}
\begin{center}
\begin{tabular}{|c|c|c|c|}\hline
& &$\mbar$ (GeV) &$\mbar$ (GeV)\\ 
\raisebox{1.5ex}[0cm][0cm]{$\beta$} &\raisebox{1.5ex}[0cm][0cm]{$a^{-1}$
(GeV)}& eq. (\ref{eq:mbar1}) & eq. (\ref{eq:mbar2})\\  
\hline 
6.0&2.0(2)&4.37(5)&4.45(5)\\ 
6.2&2.9(3)&4.38(7)&4.47(7)\\
6.4&3.8(3)&4.38(7)&4.47(8)\\ 
\hline
\end{tabular}
\end{center}
\caption{Values of $\mbar$ obtained using eqs. (\ref{eq:mbar1})
and (\ref{eq:mbar2}).}
\label{tab:mbar}\end{table}
As mentioned above, the difference of the results obtained using eqs.
(\ref{eq:mbar1}) and (\ref{eq:mbar2}) are due to our ignorance of
three-loop and higher-order perturbative corrections. This difference
is about 80--90\,MeV. The results for different values of the lattice
spacing are in remarkably good agreement. The errors on the values of
$\mbar$ in table~\ref{tab:mbar} are due to the uncertainties in the
values of the lattice spacing, in the values of ${\cal E}$ and in
$\alpha_s(M_Z)$. The dominant component in the errors is the first
one, and is a consequence of the fact that using different physical
quantities to determine the lattice spacing in quenched simulations,
leads to differences of $O(10\%)$ or so in the estimates of $a$ (which
correspond to differences of about 50\,MeV in $\mbar$). The errors in
the results for $\mbar$ in table~\ref{tab:mbar} are strongly
correlated and should not be combined in quadrature. For our best
value we take the results at $\beta=6.0$ and quote:
\begin{equation}
\mbar = 4.41\pm 0.05 \pm 0.10\ \mathrm{GeV}  \ .
\label{eq:mbarbest}\end{equation}
The first error in eq.~(\ref{eq:mbarbest}) is just that in
table~\ref{tab:mbar} and the second is an estimate of the uncertainty
in higher order perturbative corrections, as manifested in the
different values in the third and fourth columns of the table.

\subsubsection{The Unquenched Calculation}
\label{subsubsec:unquenched}
We now briefly comment on the result including fermion loops. Consider
eq.~(\ref{eq:deltaunquenched0}). In this case it is not true that most
of the fermionic two-loop contribution is reabsorbed into the $\msbar$
coupling constant. The two-loop fermionic contribution is large,
particularly with the improved action. Even with the tree-level improved
action ($c_{\mathrm{SW}}=1$) the fermionic two-loop contribution to
$\delta m$ varies from about 80 to about 150\,MeV for lattice spacings
in the range considered in this paper.  For full $O(a)$ improvement one
expects $c_{\mathrm{SW}}$, and hence the two-loop contributions, to be
larger still. Since the fermionic contribution to $\delta m$ is large,
it is likely that unquenched results for $\cal E$ will be significantly
different from those in eqs.~(\ref{eq:cale60})--(\ref{eq:cale64}) for
the corresponding values of the lattice spacing. Of course we are not
yet in a position to check whether this expectation is true.

\subsubsection{Comparison with the One-Loop Result for $\mathbf{\mbar}$}
\label{subsubsec:oneloop}
In table~\ref{tab:mbar1loop} we present the results for $\mbar$
obtained using one-loop perturbation theory in terms of
$\alpha_s(\mbar)$. The
\begin{table}
\begin{center}
\begin{tabular}{|c|c|c|c|}\hline
& &$\mbar$ (GeV) &$\mbar$ (GeV)\\ 
\raisebox{1.5ex}[0cm][0cm]{$\beta$} &\raisebox{1.5ex}[0cm][0cm]{$a^{-1}$
(GeV)}& eq. (\ref{eq:mbar1}) & eq. (\ref{eq:mbar2})\\  
\hline 
6.0&2.0(2)&4.52(4)&4.60(4)\\ 
6.2&2.9(3)&4.63(5)&4.75(5)\\
6.4&3.8(3)&4.78(6)&4.92(7)\\ 
\hline
\end{tabular}
\end{center}
\caption{Values of $\mbar$ obtained using eqs. (\ref{eq:mbar1}) and
(\ref{eq:mbar2}) in each case dropping the terms which are explicitly of
$O(\alpha_s^2(\mbar)\,)$.}
\label{tab:mbar1loop}\end{table}
errors in table~\ref{tab:mbar1loop} are also those due to the
uncertainties in the values of the lattice spacings, ${\cal E}$ and
$\alpha_s(M_Z)$.  Using this procedure, the variation of the results
with $\beta$ is relatively large and it is difficult to extract a
meaningful result for $\mbar$.

In many calculations in which the perturbative matching factors are
known only to one-loop order a different procedure is
followed. Instead of using $\alpha_s(\mbar)$ as the coupling constant
in lattice perturbation theory, one uses $\alpha_s(q^*)$ where $q^*$
is an estimate of the ``typical" loop momentum in the
process~\cite{lm}. We choose to define $q^*$ by:
\begin{equation}
\log[(aq^*)^2]=\frac{1}{X}\int \frac{d^3k}{(2\pi)^3}
\ \frac{\log[2A(k)]}{A(k)}\ ,
\end{equation}
where $X$ and $A(k)$ are defined in eqs.~(\ref{eq:xdef}) and
(\ref{eq:akdef}) respectively. With this definition $q^*a=1.446$. 
Alternatively one might use a ``boosted" lattice coupling
constant~\cite{lm}. In order to get an estimate of the precision and
reliability of such procedures we define
\begin{equation}
\varepsilon(\delta m) = (\delta m)_{\textrm{two-loops}} - 
(\delta m)_{\textrm{one-loop}} \ ,
\label{eq:epsdef}\end{equation}
where $(\delta m)_{\textrm{two-loops}}$ is the two-loop result given
in eq.~(\ref{eq:deltanew}) and $(\delta m)_{\textrm{one-loop}}$ is the
one-loop result obtained using one of the procedures described above. 
We find the following:
\begin{enumerate}
\item[i)] The one-loop result $\delta m=2.1173\,\alpha_s(q^*)\,a^{-1}$
reproduces the two-loop result in eq.~(\ref{eq:deltanew}) remarkably
well when the $\msbar$ coupling constant is used. Specifically
$\varepsilon(\delta m)\simeq$\,10, 15 and 35\,MeV for $\beta=$6.0, 6.2 and
6.4 respectively.
\item[ii)] The one-loop result varies significantly with the choice of
``reasonable" coupling. For example, using the coupling defined from the
inter-quark potential, $\alpha_V(q^*)$, one finds $\varepsilon(\delta
m)\simeq\,-210, -235$ and $-315$\,MeV  for $\beta=$6.0, 6.2 and 6.4
respectively~\footnote{For the numerical value of $\alpha_V(q^*)$ we
take $\alpha_V(q^*)=\alpha_s(q^*)(1+0.822\,\alpha_s(q^*)\,)$, and use the
physical value of the $\msbar$ coupling $\alpha_s(q^*)$ obtained
by evolving the result $\alpha_s(M_Z)=0.118(3)$ using the two-loop
renormalisation group equation.}.
\item[iii)] The use of a boosted coupling constant,
$\tilde\alpha_s=\alpha_0/u_0^4$, where $\alpha_0$ is the lattice
coupling and $u_0$ is some estimate of the average value of the
link-variable, underestimates the perturbative corrections
significantly. For example, using the trace of the plaquette to define
$u_0$, $\varepsilon(\delta m)\simeq$\, 510,  610 and 670\,MeV  for
$\beta=$6.0, 6.2 and 6.4 respectively. This is the procedure which was
used in ref.~\cite{gimenez}, and is the main reason for the low central
quoted in that paper ($\mbar=4.15\pm0.05\pm0.20$\,GeV), where the error of
200\,MeV was an estimate of the effects of higher orders in perturbation
theory.
\end{enumerate}
Thus we see that the range of one-loop perturbative results for $\delta
m$, and hence for $\mbar$ is very large; indeed it is considerably
larger than the precision required for the lattice results for the quark
mass to be phenomenologically interesting. This is the main motivation
for the two-loop calculation presented in this paper.

It should also be noted that the two-loop result in the (continuum)
perturbation series relating $m_b^{\textrm{pole}}$ and $\mbar$ is also
large (changing the result for $\mbar$ by $O(250\,$MeV$)\,)$.
Resummation techniques can also be used to try to improve the
convergence of this series (see for example refs.~\cite{resum} and
references therein).

\section{Conclusions}
\label{sec:concs}

The mass of the $b$-quark is one of the fundamental parameters of the
standard model of particle physics. It can be determined from lattice
simulations of the HQET, by computing the time behaviour of hadronic
correlation functions. An ingredient of such a determination of $\mbar$,
is the perturbation series for $\delta m$, the mass generated in the
lattice formulation of the HQET. In this paper we have evaluated the
two-loop term in the lattice perturbation series for $\delta m$, and the
result is presented in eq.~(\ref{eq:deltam2loop}). Combining our result
for $\delta m$ with the  values of $\cal E$ obtained by the APE
collaboration, we obtain the result for $\mbar$, given in
eq.~(\ref{eq:mbarresult}).

We would like to determine $\mbar$ with a precision better than
$O(\Lambda_{\textrm{QCD}})$. This is an example of a calculation of a
power correction (at leading order we can use the mass of the
pseudoscalar meson to estimate the quark mass). As with any power
correction, to achieve the required precision is very
difficult~\cite{ht}. We have argued in section~\ref{subsubsec:oneloop}
that, using one-loop matching, the uncertainty in the determination of
$\mbar$ is several hundred MeV. Our estimate (which in reality can only
be an educated ``guesstimate") of the uncertainty in $\mbar$ after the
two-loop effects are included is that it is of $O(100\,\textrm{MeV})$. 

We consider the question of the precision which can be reached when
only using one-loop matching to be so important that we repeat here
part of the discussion of section~\ref{subsubsec:oneloop}. At one-loop
level we have
\begin{equation}
\delta m = 2.1173\,\alpha_s\, a^{-1}\ .
\end{equation}
As an example let us consider the simulation at $\beta=\,6.0$ (for
which $a^{-1}=2.0(2)\,$GeV). Now as ``reasonable'' choices of
$\alpha_s$ we take~\footnote{In the following examples we take the 
central value $a^{-1}=2\,$GeV. There is a 10\% error associated
with the uncertainty in the scale, but as we wish to study the 
variation with the choice of coupling constant, we keep a fixed
value of $a$.}:
\begin{enumerate}
\item[i)] the $\msbar$ coupling at $q^*$, $\alpha_s(q^*)\simeq
0.253$, so that $\delta m\simeq 1.07\,$GeV.
\item[ii)] the coupling in the ``potential''-scheme at $q^*$, 
$\alpha_V(q^*)=0.306$, so that $\delta m\simeq 1.29\,$GeV.
\item[iii)] the ``boosted'' coupling, defined by $\alpha_0/u_0^4
\simeq 0.134$, where $u_0$ is an estimate of the avearge link-variable,
obtained from the fourth root of the plaquette. With this boosted 
coupling $\delta m\simeq 0.57\,$GeV.
\end{enumerate}
Since the perturbative coefficient is relatively large (2.1173 rather
than $1/\pi$ say) and $a^{-1}\gg\Lambda_{\textrm{QCD}}$, the spread of
results obtained using reasonable choices for the expansion parameter
$\alpha_s$ is several hundred MeV.  It is, therefore, not possible to
achieve a precision in $\delta m$ (and hence in $\mbar$) which is
better than $O(\Lambda_{\textrm{QCD}})$ without calculating the
two-loop (or even higher order) matching coefficients. This is an
example of a generic problem in the evaluation of power corrections,
and is also not restricted to lattice computations.

It should be said that our view that the ignorance of higher order
perturbative coefficients implies that the uncertainties in the
results for power corrections to physical quantities are large is not
universally accepted. In this paper we have confirmed our view with a
specific two-loop calculation. For $m_b$ the leading correction is
linear, and correspondingly we have had to subtract the linear
divergence (i.e. the terms which diverged linearly in $1/a$) in $\cal
E$ using perturbation theory. In other important examples one needs to
subtract quadratic or even higher order divergences and the difficulty
to achieve the required precision increases enormously.

\subsection*{Acknowledgements} We warmly thank Rajan Gupta and
the participants of the International Workshop on Perturbative and
Non-Perturbative Aspects of the Standard Model, Santa Fe, August 1998,
for lively discussions on the material of this paper.

GM thanks Alvaro de Rujula and the Theory Division at CERN for
hospitality while this work was completed and acknowledges partial
support from MURST. CTS acknowledges partial support from PPARC
through grants GR/L56329 and PPA/G/S/1997/00191.

\newpage
%\begin{center}
%\Large{\textbf{Appendix}}
%\end{center}
%\vspace{-25pt}
\appendix
\numberwithin{equation}{section}
\section{Appendix: Two-Loop Contribution to Large Wilson Loops}
Our calculation of $\delta m$ is based on the fact that it is
proportional to the logarithm of the perimeter term in the
perturbative expansion of large Wilson loops. In this appendix we
present the details of the evaluation of the two-loop contribution to
large Wilson loops. 

The appendix is organised as follows. In the opening section we define
the multidimensional integrals which appear in our result for $W_4$
and present their numerical values. A description of the evaluation of
the terms in $W_4$ which grow at least linearly with $R$ and $T$ in
the pure gauge-theory (i.e. in the theory without light quarks) is
presented in section~\ref{subsec:w4gluon}. The fermionic contributions
to $W_4$ are evaluated in section~\ref{subsec:w4fermion}.

\subsection{Integrals}\label{subsec:integrals}
The two-loop contribution to $\delta m$ will be presented in terms of
a number of multidimensional integrals,
$X,L,Y,W,V_{1,2},V_{\textrm{W}}$ and $V_{\textrm{SW}}$, which have to
be evaluated numerically. In this subsection we define these integrals
and present their values. All the integrals have limits of integration
from --$\pi$ to $\pi$ for each momentum component.
\begin{enumerate}
\item[\textbf{1.}] The first three integrals are one-loop ones. In each
case it is straightforward to perform the integral over one component
of momentum analytically, leaving a three-dimensional integral to be 
performed numerically. The first of these is
\begin{equation}
X\equiv\int \frac{d^3k}{(2\pi)^3}\frac{1}{A(k)}=0.50546
\label{eq:xdef}\end{equation}
where
\begin{equation}
A(k)=\sum_{i=1}^3 (1-\cos(k_i))\ .\label{eq:akdef}
\end{equation}
$X$ is the integral which contibutes to the one-loop component of
$\delta m$.
\item[\textbf{2.}] The second one-loop integral is
\begin{equation}
L\equiv\int \frac{d^3k}{(2\pi)^3}\frac{1}{\sqrt{(1+A(k))^2-1)}}=0.30987\ .
\label{eq:ldef}\end{equation}
\item[\textbf{3.}]The third integral arises in the calculation of the 
wave-function renormalisation at one-loop level:
\begin{equation}
Y(T)\equiv\int\frac{d^3 k}{(2\pi)^3}\frac{1}{A\sqrt{(1+A)^2-1}}(1-\beta^T),
\label{eq:ydef}\end{equation}
where
\begin{equation}
\beta=1+A-\sqrt{(1+A)^2-1}\ ,
\end{equation}
and $A$ and $\beta$ are implicitly functions of $k$. Without the
$\beta^T$ term on the right-hand side of eq.~(\ref{eq:ydef}) the
integral would be divergent (it depends logarithmically on T). In the
final result for $\delta m$, any dependence on $Y(T)$ or $Y(R)$ must
cancel.
\item[\textbf{4.}] The remaining integrals are 7-dimensional and for these the
results are generally obtained with poorer relative precision.
The first of these
is
\begin{eqnarray}
W&\equiv&\int\frac{d^3k}{(2\pi)^3}\frac{1}{A(k)}
\int\frac{d^3p}{(2\pi)^3}\frac{1}{A(p)}\nonumber\\ 
& & \hspace{-0.6in}\cdot\int\frac{dq}{2\pi}\frac{\sin^2(q)}
{(4\sin^2(q/2)+2A(k))(4\sin^2(q/2)+2A(p))}=1.313(3)\cdot 10^{-2}\ .
\label{eq:wdef}\end{eqnarray}
\item[\textbf{5.}] There are two integrals which arise from the gluonic
contribution to the vacuum-polarisation. The first of these is
\begin{eqnarray}
V_1&\equiv& \int\frac{d^3p}{(2\pi)^3}
\frac{d^4k}{(2\pi)^4}
\frac{(1+\cos(k_\mu))}
{A^2(p)(4-C_4(k))}
\left\{\frac{4-C_4(2p+k)}{4-C_4(p+k)}
-1\right\}
\nonumber\\ 
& = & 7.008(5)\cdot 10^{-2}\ ,
\label{eq:v1def}\end{eqnarray}
where $p_\mu=0$ ($\mu$ is a fixed direction and the integral over $p$
is over the 3 components other than the $\mu$-component). The
function $C_4$ is defined by
\begin{equation}
C_4(r)\equiv\sum_{\lambda=1}^4 \cos (r_\lambda)\ .
\end{equation}
\item[\textbf{6.}] The second integral arising from the gluonic
vacuum-polarisation graphs is
\begin{eqnarray}
V_2&\equiv& \int\frac{d^3p}{(2\pi)^3}
\frac{d^4k}{(2\pi)^4}
\,\frac{\sin^2(k_\mu)}
{A^2(p)(4-C_4(k))}\,
\,\left\{\frac{1+\sum_{i\neq\mu}\cos(p_i)}{2\,(4-C_4(p+k)\,)}
-\frac{2}{4-C_4(k)}\right\}
\nonumber\\ 
&=&-2.130(1)\cdot 10^{-2}\ ,
\label{eq:v2def}\end{eqnarray}
where, again, $\mu$ is a fixed direction and $p_\mu=0$.
\item[\textbf{7.}] The final two integrals correspond to graphs containing
fermionic contributions to the vacuum-polarisation. For Wilson-fermions
the relevant integral is
\begin{equation}
V_{\mathrm{W}}\equiv \int\frac{d^3p}{(2\pi)^3}\frac{d^4k}{(2\pi)^4}
\frac{1}{A^2(p)}\,\frac{1}{S(k)}\,\left[\frac{Z_{\mathrm{W}}(p,\,k)}{S(q)}
-\frac{Z_{\mathrm{W}}(p=0,k)}{S(k)}\right]\ ,
\label{eq:vWdef}\end{equation}
where $q\equiv p+k$, $p_\mu=0$, 
\begin{equation}
S(k)=\sum_{\lambda=1}^4\sin^2(k_\lambda) + W^2(k)\ ,
\label{eq:sdef}\end{equation}
\begin{equation}
W(k)=2\sum_{\lambda=1}^4 \sin^2\left(\frac{k_\lambda}{2}\right)\ ,
\label{eq:wkdef}\end{equation}
and
\begin{eqnarray}
Z_{\mathrm{W}}(p,k) & = & -8 \cos^2(k_\mu)\sin^2(k_\mu)
+4\sum_{\lambda=1}^4\sin(k_\lambda)\sin(q_\lambda)
\nonumber\\
&&\hspace{-0.4in}+4\cos(2k_\mu)W(k)W(q)
-8\cos(k_\mu)\sin^2(k_\mu)[W(k)+W(q)]\ .
\label{eq:zWdef}\end{eqnarray}
$W(k)$, the Wilson term in the fermion propagator defined in
eq.~(\ref{eq:wkdef}), should not be confused with the integral $W$
defined in eq.~(\ref{eq:wdef}).
We find that
\begin{equation}
V_{\mathrm{W}} = -2.145(1)\cdot10^{-2}\ .
\label{eq:vwdef}\end{equation}
\item[\textbf{8.}]
Finally we present the integral, $V_{\mathrm{SW}}$ required to evaluate
the additional fermionic contributions when the SW-improved action is
used,
\begin{equation}
V_{\mathrm{SW}}\equiv \int\frac{d^3p}{(2\pi)^3}\frac{d^4k}{(2\pi)^4}
\frac{1}{A^2(p)}\,\frac{Z_{\mathrm{SW}}}{S(k)S(q)}\ ,
\label{eq:vswdef}\end{equation}
where, again, $q=p+k$, $p_\mu=0$ and
\begin{eqnarray}
Z_{\mathrm{SW}}&=&c_{\mathrm SW}\,\left\{
4\sin^2(k_\mu)\sum_{\lambda=1}^4
\sin(p_\lambda)\,[\sin(q_\lambda)-\sin(k_\lambda)]\right.
\nonumber\\
&&
\hspace{-0.3in}\left. -4\cos(k_\mu)\,\left(
W(k)\sum_{\lambda=1}^4 \sin(p_\lambda)\sin(q_\lambda)
-W(q)\sum_{\lambda=1}^4 \sin(p_\lambda)\sin(k_\lambda)\right)\right\}
\nonumber\\
&&\hspace{-0.6in}
-c_{\mathrm SW}^2\,\left\{
\left(2\sin^2(k_\mu)-\sum_{\lambda=1}^4 \sin(q_\lambda)\sin(k_\lambda)
+W(k)W(q)\right)\sum_{\rho=1}^3\sin^2(p_\rho)\right.\nonumber\\ 
& & \left.
+2 \sum_{\lambda=1}^4 \sin(p_\lambda)\sin(k_\lambda)
\sum_{\sigma=1}^4 \sin(p_\sigma)\sin(q_\sigma)\right\}\ .
\label{eq:zswdef}\end{eqnarray}
Evaluating the integrals numerically we find
\begin{equation}
V_{\mathrm{SW}} = 2.63(2)\cdot 10^{-3}\,c_{\mathrm{SW}}
- 2.974(1)\cdot10^{-2}\,\,c_{\mathrm{SW}}^2\ . 
\end{equation}
\end{enumerate}

\subsection{The Gluonic Contribution to $\mathbf{W_4(R,T)}$}
\label{subsec:w4gluon}

In this section we outline the evaluation of the gluonic contribution to
the two-loop term in the Wilson loop, $-g^4\,W_4(R,T)$.
Following ref.~\cite{hk}, we distinguish 5 contributions to 
$W_4$~\footnote{The five contributions in eq.~(\ref{eq:fivec}) are
defined explicitly in eqs.~(3.6)--(3.10) of ref.~\cite{hk}.}:
\begin{equation}
W_4(R,T)=W_{S_1}+W_I+W_{II}+W_{VP}+\overline W_{VP}\ .
\label{eq:fivec}\end{equation}
We now evaluate each of these in turn, picking up the contributions
which grow as $R,\ T$, $R^2$, $T^2$ or $RT$.

\paragraph{$\mathbf{W_{S_1}}$:}
This contribution comes from the ``spider" graph and is given in
eq.~(3.6) of ref.~\cite{hk}. It is proportional to
\begin{multline}
\int\frac{d^4k}{(2\pi)^4}\frac{d^4p}{(2\pi)^4}\,
\frac{1}{D(p)D(k)D(p+k)}\Bigg[\Big\{\frac{\sin^2(\frac12 p_\mu R)}
{\sin(\frac12 p_\mu)}\sin(\half p_\nu T)\cos(\half p_\nu)
\sin(\half (2k+p)_\mu)\\ 
\times\Big[\frac{\sin(\frac12 p_\nu T)\sin(\frac12 (2k+p)_\nu)}
{\sin(\frac12 p_\nu)\sin(\frac12 k_\nu)\sin(\frac12 (p+k)_\nu)}-
\frac{\sin(\frac12 (2k+p)_\nu T)}{\sin(\frac12 k_\nu)\sin(\frac12(p+k)_\nu)}
\Big]\\
+4\frac{\sin(\frac12 p_\mu R)}{\sin(\frac12 p_\mu)}
\frac{\sin(\frac12 k_\mu R)}{\sin(\frac12 k_\mu)}
\frac{\sin(\frac12 (p+k)_\nu T)}{\sin(\frac12 (p+k)_\nu)}
\sin(\half p_\nu T)\cos(\half k_\nu T)\\ 
\times\cos(\half(p+k)_\mu R)
\cos(\half(p+k)_\mu)\sin(\half((k-p)_\nu))\Big\}+\{(\mu,R)
\leftrightarrow(\nu,T)\}\Bigg]\ ,
\end{multline}
where $D(p)$ is defined in eq.~(\ref{eq:ddef}).
By detailed inspection of the infrared behaviour of the integral,
one can see that there is insufficient enhancement
from the denominators to give a term 
proportional to $R$ or $T$. Thus there is no contribution to $\delta m$
from $W_{S_1}$.

\paragraph{$\mathbf{W_{I}}$:}
$W_I$ is defined to be the ``expectation value of the non-abelian part
of the order $g^4$ term in the expansion of the Wilson
loop"~\cite{hk}.  It is presented in eq.~(3.7) of ref.~\cite{hk} and
contains several terms which we consider in turn. The first five
contributions, $T_1$--$T_5$, are products of the one-loop integrals
defined in section~\ref{subsec:integrals} and are hence relatively
straightforward to evaluate. We therefore simply define these
contributions and present the corresponding results in terms of the
integrals $X$, $L$, and $Y$.

\begin{eqnarray}
T_1& =& \hspace{-8pt}-\frac{(N^2-1)}{3} \Delta_0 
\Bigg[\Bigg\{\int\frac{d^4p}{(2\pi)^4}
\frac{\sin^2(\frac12p_\nu T)}{D(p)}
\Big(\frac{\sin^2(\frac12 p_\mu R)}{\sin^2(\frac12 p_\mu)}
+\frac12\frac{\sin(\frac12 p_\mu R)\sin(\frac12 p_\mu (R-2))}
{\sin^2(\frac12 p_\mu)}\Big)\hspace{-3pt}\Bigg\}\nonumber\\
&&\hspace{1.5in}+\{(\mu,R)\leftrightarrow(\nu,T)\}\Bigg]
\label{eq:t1def}\\ 
&=&-\frac{(N^2-1)}{16}(R+T)XL\ .\label{eq:T1res}
\end{eqnarray}
In eq.~(\ref{eq:t1def})
\begin{equation}
\Delta_0=\int\frac{d^4k}{(2\pi)^4}\,\frac{1}{D(k)}\ =\ \frac{L}{2}\ .
\label{eq:delta0def}\end{equation}
We repeat that we only keep terms which grow at least linearly with $R$
and $T$.

\begin{eqnarray}
T_2&=&-\frac{(N^2-1)}{2}\Bigg[\Bigg\{\int\frac{d^4k}{(2\pi)^4}
\frac{1}{D(k)}\Big(\frac{\sin^2(\frac12 k_\nu T)}
{\sin^2(\frac12 k_\nu)}+\frac16 
\frac{\sin^2(\frac12 k_\mu R)}
{\sin^2(\frac12 k_\mu)}\Big)\nonumber\\
& & \times
\int\frac{d^4p}{(2\pi)^4}\frac{\sin^2(\frac12 p_\nu T)}
{D(p)}\frac{\sin^2(\frac12 p_\mu R)}
{\sin^2(\frac12 p_\mu)}\Bigg\} 
+\{(\mu,R)\leftrightarrow(\nu,T)\}\Bigg] \\
&\simeq&
-\frac{(N^2-1)}{8}[XT-Y(T)]\,[XR-Y(R)]- \nonumber\\
& & \hspace{0.3in}\frac{(N^2-1)}{96}[XR-Y(R)]^2
- \frac{(N^2-1)}{96}[XT-Y(T)]^2\ .\label{eq:T2res}   
\end{eqnarray}

%\noindent\textbf{3:}
\begin{eqnarray}
T_3&=&\frac{(N^2-1)}{12}\Bigg[\Big\{
\Big[\int\frac{d^4p}{(2\pi)^4}
\frac{\sin^2(\frac12 p_\nu T)}{D(p)}
\frac{\sin^2(\frac12 p_\mu R)}
{\sin^2(\frac12 p_\mu)}\Big]^2\Big\}
+\{(\mu,R)\leftrightarrow(\nu,T)\}\Bigg] \\
&\simeq & \frac{N^2-1}{192}\left\{[XR-Y(R)]^2 + [XT-Y(T)]^2\right\}\ .   
\end{eqnarray}

%\noindent\textbf{4:}
\begin{eqnarray}
T_4&=&\frac{2(N^2-1)}{3}
\int\frac{d^4k}{(2\pi)^4}
\frac{\sin^2(\frac12 k_\nu T)}{D(k)}
\frac{\sin^2(\frac12 k_\mu R)}
{\sin^2(\frac12 k_\mu)}\nonumber\\ 
& & \hspace{0.2in}\times\int\frac{d^4p}{(2\pi)^4}
\frac{\sin^2(\frac12 p_\mu R)}{D(p)}
\frac{\sin^2(\frac12 p_\nu T)}
{\sin^2(\frac12 p_\nu)}\\
& \simeq & \frac{(N^2-1)}{24}(XR-Y(R))\,(XT-Y(T))\ .
\end{eqnarray}

\begin{eqnarray}
T_5&=&\frac{(N^2-1)}{4}
\int\frac{d^4k}{(2\pi)^4}
\frac{\sin^2(\frac12 k_\mu R)}
{D(k)\sin^2(\frac12 k_\mu)}
\int\frac{d^4p}{(2\pi)^4}
\frac{\sin^2(\frac12 p_\nu T)}
{D(p)\sin^2(\frac12 p_\nu)}\\
&\simeq & \frac{(N^2-1)}{16}(XR-Y(R))\,(XT-Y(T))\ .
\end{eqnarray}

The remaining three terms contain nested integrals and are
considerably more complicated to evaluate. The first of these
is
\begin{multline}
T_6=\frac{(N^2-1)}{32}
\iint\frac{d^4k}{(2\pi)^4}\frac{d^4p}{(2\pi)^4}
\frac{1}{D(p)D(k)}\\ 
\Bigg[\Bigg\{\Big[
\frac{\sin(\frac12(p+k)_\mu R)\sin(\frac12(p-k)_\mu)-
\sin(\frac12(p-k)_\mu R)\sin(\frac12(p+k)_\mu)}
{\sin(\frac12(p+k)_\mu)}\Big]^2\\ 
\times\frac{\sin^2(\frac12(p+k)_\nu T)}
{\sin^2(\frac12 p_\mu)\sin^2(\frac12 k_\mu)}\Bigg\}
+\{(\mu,R)\leftrightarrow(\nu,T)\}\Bigg] \ .\hspace{1in}
\label{eq:t6def}\end{multline}
In order to extract the behaviour of this integral at large $R$ and $T$
consider the following two-dimensional integral 
\begin{equation}
I=\int\frac{dk_\mu}{2\pi}\frac{dp_\mu}{2\pi}\frac{1}{D(p)D(k)}
\frac{N}{4\sin^2(p_\mu/2)\,4\sin^2(k_\mu/2)\,4\sin^2((p+k)_\mu/2)}\ ,
\label{eq:idef}\end{equation}
where
\begin{equation}
N=\left[4\sin\left(\frac{(p+k)_\mu R}{2}\right)\,\sin\left(\frac{(p-k)_\mu}{2}\right)-
4\sin\left(\frac{(p-k)_\mu R}{2}\right)
\sin\left(\frac{(p+k)_\mu}{2}\right)\right]^2\ .
\end{equation}
It is convenient to use partial fractions to simplify I:
\begin{equation}
I=I_1-(I_2+I_3)+I_4\ ,
\end{equation}
where
\begin{eqnarray*}
I_1 & =& \frac{1}{4A(p)A(k)}\int\frac{dp_\mu}{2\pi}\frac{dk_\mu}{2\pi}
\frac{N}{4\sin^2(p_\mu/2)\,4\sin^2(k_\mu/2)\,4\sin^2((p+k)_\mu/2)}\\ 
I_2 & =& \frac{1}{4A(p)A(k)}\int\frac{dp_\mu}{2\pi}\frac{dk_\mu}{2\pi}
\frac{N}{4\sin^2(p_\mu/2)\,(4\sin^2(k_\mu/2)+2A(k))\,4\sin^2((p+k)_\mu/2)}\\ 
I_3 & =& \frac{1}{4A(p)A(k)}\int\frac{dp_\mu}{2\pi}\frac{dk_\mu}{2\pi}
\frac{N}{(4\sin^2(p_\mu/2)+2A(p))\,4\sin^2(k_\mu/2)\,4\sin^2((p+k)_\mu/2)}\\ 
I_4 & =& \frac{1}{4A(p)A(k)}\int\frac{dp_\mu}{2\pi}\frac{dk_\mu}{2\pi}
\frac{N}{4\sin^2((p+k)_\mu/2)}\\
& & \hspace{1.5in}\times\frac{1}{(4\sin^2(p_\mu/2)+2A(p))\,(4\sin^2(k_\mu/2)+2A(k))}
\ .\end{eqnarray*}
$A(p)$ and $A(k)$ are defined in eq.~(\ref{eq:akdef}), where the sum is over the
three components of $p$ and $k$ other than the $\mu$-component.

$I_1$--$I_4$ can be evaluated by changing integration variables to
$z=\exp(i p_\mu)$ and $\omega=\exp(i (p+k)_\mu)$ and using Cauchy's contour
integration theory. In terms of these complex variables
\begin{equation}
N=\omega^{-(R+1)}\left[(1+z)(\omega^R-1)(1-\omega/z)
-(\omega-1)(z^R+1)(1-(\omega/z)^R)\right]^2\ .
\end{equation}
In this way we obtain
\begin{eqnarray*}
I_1&=&\frac{R(R-1)}{4}\,X^2\\
I_2&=&\frac{3R}{4}XY(R)-\frac{R}{4}X^2+\frac{R}{4}XL\\
I_3&=&I_2\\
I_4&=&RW\ .
\end{eqnarray*}
Note that
\begin{equation}
\int\frac{dp_\mu}{2\pi}\frac{dk_\mu}{2\pi}\frac{N}{64\sin^2(p_\mu/2)\,\sin^2(k_\mu/2)\,
\sin^2((p+k)_\mu/2)}=R(R-1)\ .
\end{equation}
Since we are only interested in identifying the terms which grow at least linearly
with $R$ or $T$, we can replace $\sin^2(\half(p+k)_\nu T)$ in eq.~(\ref{eq:t6def})
with the factor $\half$, thus obtaining
\begin{equation}
T_6=\frac{N^2-1}{16}\left[\frac{R(R+1)}{4}X^2-\frac{3R}{2}XY(R)
-\frac{R}{2}XL+RW\right]
+ R\leftrightarrow T\ .
\end{equation}

The second nested integral is 
\begin{multline}
T_7 = \frac{N^2-1}{6}
\iint\frac{d^4k}{(2\pi)^4}\frac{d^4p}{(2\pi)^4}
\frac{1}{D(p)D(k)}\\
\Big[\Big\{\sin^2(\half p_\nu T)
\frac{\sin(\frac12p_\mu R)}{\sin(\frac12 p_\mu)}
\frac{\sin(\frac12k_\mu R)}{\sin(\frac12 k_\mu)}
\frac{\sin(\frac12(p+k)_\mu R)}{\sin(\frac12 (p+k)_\mu)}
\Big\}
+\{(\mu,R)\leftrightarrow(\nu,T)\}\Big] \ .
\end{multline}
Using the fact that
\begin{equation}
\int\frac{dp_\mu}{2\pi}\int\frac{dk_\mu}{2\pi}\frac{\sin(p_\mu R/2)
\sin(k_\mu R/2)\sin((p+k)_\mu R/2)}
{\sin(p_\mu/2)\sin(k_\mu/2)\sin((p+k)_\mu/2)}=R\ ,
\end{equation}
we readily obtain
\begin{equation}
T_7=\frac{N^2-1}{48}X^2(R+T)\ .
\end{equation}

The final contribution to $W_I$ is
\begin{multline}
T_8=-\frac{(N^2-1)}{6}
\iint\frac{d^4k}{(2\pi)^4}\frac{d^4p}{(2\pi)^4}
\frac{1}{D(p)D(k)}\\ 
\Bigg[\Bigg\{\sin^2(\half p_\nu T)
\frac{\sin(\frac12p_\mu R)}{\sin(\frac12 p_\mu)}
\Big[\frac{\sin(\frac12p_\mu(R-2))\cos((p+k)_\mu)}
{\sin(\frac12 p_\mu)\sin^2(\frac12 k_\mu)}\hspace{0.5in}\\ 
-\frac{\sin(\frac12(p+k)_\mu(R-2))\cos(\frac12k_\mu(R+1)+p_\mu)}
{\sin(\frac12 (p+k)_\mu)\sin^2(\frac12 k_\mu)}
+\frac{\sin(\frac12 p_\mu(R-2))\cos(\frac12p_\mu-k_\mu)}
{\sin(\frac12 p_\mu)\sin(\frac12 k_\mu)\sin(\frac12(p+k)_\mu)}\\ 
-\frac{\sin(\frac12 k_\mu(R-2))
\cos(\frac12(p+k)_\mu R+\frac12k_\mu-p_\mu)}
{\sin(\frac12 (p+k)_\mu)\sin^2(\frac12 k_\mu)}\\ 
+\frac{\sin(\frac12(p+k)_\mu(R-2))
\cos(\frac12k_\mu(R+2)+\frac12p_\mu)}
{\sin(\frac12 k_\mu)\sin(\frac12 p_\mu)\sin(\frac12 (p+k)_\mu)}\\
-\frac{\sin(\frac12 k_\mu(R-2))
\cos(\frac12(p-k)_\mu R-k_\mu-p_\mu)}
{\sin(\frac12 p_\mu)\sin^2(\frac12 k_\mu)}\Big]\Bigg\}
+\{(\mu,R)\leftrightarrow(\nu,T)\}\Bigg] \ ,
\end{multline}
for which, by using partial fractions as in the evaluation of $T_6$, 
we find
\begin{equation}
T_8=\frac{(N^2-1)}{48}X\left[-XR^2-XR+5Y(R)R+4LR\right] +
R\leftrightarrow T\ .
\end{equation}

Summing up the terms $T_1$ to $T_8$ to obtain $W_I$ we find
\begin{multline}
W_I=\sum_{i=1}^8 T_i  = 
-\frac{(N^2-1)}{48}(XR-Y(R))(XT-Y(T))\\  
 - \frac{(N^2-1)}{192}\left\{(XR-Y(R))^2 + (XT-Y(T))^2\right\}\\
 +  \frac{(N^2-1)}{192}\left\{-R^2X^2+3RX^2+2XY(R)R 
 -2RXL +12RW + R\leftrightarrow T\right\} \ .
\label{eq:wires}\end{multline}

\paragraph{$\mathbf{W_{II}}$:}
$W_{II}$ is the ``abelian part of the expansion of order $g^4$", and
is defined in eq.~(3.8) of ref.~\cite{hk}.
It is very straightforward to evaluate:
\begin{eqnarray}
W_{II}&=&-\frac{(2N^2-3)(N^2-1)}{6N^2}\,\left[\overline{W}_2(R,T)\right]^2\\ 
& = & -\frac{(2N^2-3)(N^2-1)}{96N^2}\,\left[X(R+T)-Y(R)-Y(T)\right]^2\ ,
\label{eq:wiires}\end{eqnarray}
where $\overline{W}_2$ has been defined in eq.~(\ref{eq:w2res}).

\paragraph{$\mathbf{W_{\textrm{VP}}}$:}
$W_{\textrm{VP}}$ is part of the contribution to the diagrams containing
gluonic contributions to the vacuum polarisation, and is defined
in eq.(3.9) of ref.~\cite{hk}.
It can be written in the form:
\begin{equation}
W_{\textrm{VP}}=\frac{(N^2-1)}{N}\,\frac{(R+T)}{8}\int\frac{d^3p}{(2\pi)^3}
\left(
\frac{\Pi^a_{\mu\mu}(p;p_\mu=0)-\Pi^a_{\mu\mu}(p=0)}{A^2(p)}\right)\ ,
\label{eq:wvp}\end{equation}
where there is no implied sum over $\mu$, and
\begin{multline}
\Pi_{\mu\mu}^a(p;p_\mu=0) = \frac{N}{6}\big(\frac{7L}{2}-\frac58\big)A(p)+
\\
N\int\hspace{-5pt}\frac{d^4k}{(2\pi)^4}
\frac{1}{D(p+k)D(k)}\Big\{(1+\cos(k_\mu))(4-C_4(2p+k))
+\sin^2(k_\mu)\big(1+\sum_{i\neq\mu}\cos(p_i)\big)\Big\}\, .
\label{eq:pimumu}\end{multline}

Although $\Pi_{\mu\mu}^a(p=0)=0$, it is nevertheless convenient to
subtract it (in the form of eq.~(\ref{eq:pimumu}) with $p=0$) in the
numerator of the integrand in eq.~(\ref{eq:wvp}) and to write
$W_{\textrm{VP}}$ in terms of the integrals $V_1$ and $V_2$ defined
in eqs.~(\ref{eq:v1def}) and (\ref{eq:v2def}). This
subtraction introduces non-zero contributions to each of $V_1$ and
$V_2$, rendering these integrals convergent, but of course these
contributions cancel in $W_{\textrm{VP}}$. In this way we obtain
\begin{equation}
W_{VP}= \frac{(N^2-1)}{32}(R+T)\,\big(V_1 + 2V_2\big)
+\frac{(N^2-1)}{48}(R+T)\left(\frac{7L}{2}-\frac58\right)X\ .
\label{eq:wvpres}\end{equation}

\paragraph{$\mathbf{\overline{W}_{\textrm{VP}}}$:}

$\overline{W}_{\textrm{VP}}$, the remaining contribution to the
diagrams containing gluonic contributions to the vacuum
polarisation~\cite{hk} is defined in eq.~(3.10) of ref.~\cite{hk} and
is simple to evaluate:
\begin{equation}
\overline W_{VP} = \frac{(2N^2-3)(N^2-1)}{96N^2}\,
\left[(R+T)X-Y(R)-Y(T)\right]\ .
\label{eq:wbarvpres}\end{equation}

\subsubsection{Total Gluonic Contribution to $\mathbf{W_4}$:}
Summing up the contributions from eqs.~(\ref{eq:wires}),
(\ref{eq:wiires}), (\ref{eq:wvpres}) and (\ref{eq:wbarvpres}),
and keeping only those terms which grow at least linearly with
$R$ and $T$,
we obtain for the total gluonic contribution to $W_4$:
\begin{multline}
W_4^{\mathrm{gluon}}=-\frac{(N^2-1)^2}{32N^2}X^2(R+T)^2 
+\frac{(N^2-1)^2}{16N^2}X(R+T)(Y(R)+Y(T))\\
+\frac{(N^2-1)}{192}\left\{3X^2 - 2XL+ 12W\right\}\,(R+T)\\ 
+\frac{(N^2-1)}{96}\left\{3V_1 + 6V_2 + 7LX -
\frac{5}{4}X\right\}\,(R+T) 
+ \frac{(N^2-1)(2N^2-3)}{96N^2}X(R+T)\ .
\label{eq:w4gluonres}\end{multline}

\subsection{Evaluation of the Fermionic Contribution to $\mathbf{W_4(R,T)}$}
\label{subsec:w4fermion}
In this section we evaluate the diagrams containing fermionic
contributions to the vacuum polarisation. This contribution is also given
by eq.~(\ref{eq:wvp}) but with $\Pi^a_{\mu\mu}\rightarrow
\Pi^{\mathrm{fermion}}_{\mu\mu}$, where $\Pi^{\mathrm{fermion}}_{\mu\mu}$
is the fermionic contribution to the vacuum polarisation graph:
\begin{equation}
\Pi^{\mathrm{fermion}}_{\mu\mu}(p;p_\mu=0)=\frac{N_f}{2}
\int \frac{d^4k}{(2\pi)^4}\ \frac{Z_{\mathrm{W}}(p,k)+Z_{\mathrm{SW}}(p,k)}
{S(k)S(q)}\ ,
\end{equation}
where $q=p+k$, $S(k)$ is defined in eq.~(\ref{eq:sdef}) and
$Z_{\mathrm{W,SW}}(p,k)$ are defined in eqs.~(\ref{eq:zWdef}) and
(\ref{eq:zswdef}) respectively. $N_f$ is the number of active
light-quark flavours and we distinguish between the terms from the
Wilson Fermion action (labelled by $\scriptstyle{\mathrm{W}}$) and
the additional terms in the improved action with coefficient
$c_{\mathrm{SW}}$ (labelled by $\scriptstyle{\mathrm{SW}}$). Thus the
fermionic contribution to $W_4$ is
\begin{equation}
W_4^{\mathrm{fermion}}=\frac{(N^2-1)N_f}{16N}\,(R+T)\,\big(V_{\mathrm{W}}+
V_{\mathrm{SW}}\big)\ .
\label{eq:w4fermionres}\end{equation}

\end{document}